# Langmuir wave filamentation instability

Harvey A. Rose and L. Yin
Los Alamos National Laboratory

A Langmuir wave (LW) model is constructed whose equilibria are consistent with stimulated Raman scatter optimization, with Hamiltonian dynamics and with rotational invariance. Linear instability analysis includes terms to all orders in wave amplitude and fluctuation wavenumber expansions, $\delta k$. Resultant LW modulational instability is nonstandard: as the LW amplitude increases, unstable $\delta k$ range first expands and then shrinks to zero. Large amplitude wave model dynamics requires hyper-diffraction terms if $k\lambda_D < 0.45$, lest artificially small length scales become unstable.

# I. INTRODUCTION

High intensity laser beams are spatially smoothed[1] to avoid large scale, non-reproducible spatial variation, and attendant self-focusing. Ironically, small-scale intensity fluctuations, speckles, are thereby created, whose probability distribution has a high intensity tail, allowing stimulated Raman backscatter (SRS) in individual speckles even though the average beam intensity is below threshold. The daughter Langmuir wave (LW) is spatially localized in speckles, since that is where the LW source, which is proportional to the high frequency beat ponderomotive potential of the laser beam and scattered light, is a local maximum. Speckles are highly elongated along the laser beam: they have linear dimensions proportional to the basic, optic induced, correlation length scales: $l_\perp = F\lambda_0/\pi$, perpendicular to the beam, where, $\lambda_0$ is the laser wavelength and $F$ the optic f/#, and $l_\parallel = k_0 l_\perp^2 = 2F^2\lambda_0/\pi$, parallel to the beam. The actual ratio of speckle length (full width at half intensity max) to width is $O(10F)$. Also, deep in the kinetic regime, $e.g.$, $k\lambda_D \approx 0.5$, with $k$ the LW wavenumber, $l_\perp$ is large compared to $1/k$,

$$kl_\perp \approx F\frac{c}{v_e}\frac{\omega_{pe}}{\omega_0} = F\frac{22.6}{\sqrt{T_{keV}}}\frac{\omega_{pe}}{\omega_0}. \tag{1}$$

$v_e$ is the electron thermal speed, $T_{keV} = T_e/keV$, $T_e$ the electron temperature and $\omega_{pe}$ $(\omega_0)$ the electron plasma (laser) frequency. For example, if $T_{keV} = 4$, $\omega_{pe}/\omega_0 = 0.25$ and $F = 8$, then $kl_\perp \approx 20$. At SRS onset[2] the most intense speckles are dominant[3] and the average laser intensity, $\langle I \rangle$, is such that a speckle with intensity $I/\langle I \rangle >> 1$ has SRS power gain exponent roughly $I/\langle I \rangle$. These speckles are rare and hence spatially well separated. Current laboratory experiments in long scale length plasma have $I_{max}/\langle I \rangle = O(10)$, while for the previous example, $kl_{speckle}/2\pi \approx 600$. Hence the SRS power gain over a LW wavelength, even in the most intense speckle, is only order $\exp(1/60)$. These numerical estimates suggest that the study of nonlinear kinetic effects in a LW wavetrain is a useful starting point for modeling SRS near its onset intensity. In one spatial dimension (1D) we will show when this wavetrain is close to a BGK mode[4].

Kinetic effects include the well-known[5 6 7] trapped electron LW frequency shift, $\Delta\omega < 0$. Once $|\Delta\omega|$ exceeds that due to LW ponderomotive density depletion, which occurs for[8] $k\lambda_D \gtrsim 0.2$, we expect that trapped electron LW self-localization effects are generally[9] dominant. Other nonlinear effects that follow from trapping may be realized in 1D, such as reduction of Landau damping, while LW self-focusing, require $D \geq 2$, the main emphasis of this work.

## II. BGK MODES, OPTIMAL SRS THEORY and NON-PERTURBATIVE MODULATIONAL INSTABLILITY

No reduced model of the SRS daughter Langmuir wave can faithfully represent all physically significant features in an arbitrary SRS regime. While any particular reference regime is to some degree arbitrary, we choose a family of models that yield accurate results when Langmuir wave parameters vary slowly, because it is the simplest regime consistent with the basic premise that reduced LW models exclude rapid $1/\omega_{pe}$ $(1/k)$ time (space) scale variation and because the slowly varying regime is of intrinsic interest to SRS predictive capability: results[10] show that for SRS growth rates small compared to $\omega_{pe}$, and plasma density scale lengths large compared to a speckle length, peak speckle SRS reflectivity occurs while the LW grows slowly and coherently over many $1/\omega_{pe}$, and is spatially coherent over many LW wavelengths.

We also require that the model's small LW amplitude limit, $\phi \to 0$, recovers linear SRS theory. If this were not the case, then on the far side of the laser speckle, away from the laser, where LW fluctuations are small in the convective SRS regime, the model would fail. Since the convective gain rate is then proportional to $-\mathrm{Im}\left(1/\varepsilon_0\right)$, with $\varepsilon_0$ the linear dielectric function, a model must be aware of the plasma background distribution function, $f_0(v)$. *A fortiori,* since an isolated laser speckle cannot affect the distribution of incoming electrons, knowledge of $f_0(v)$ must be preserved even for large $\phi$ and long evolution time.

As in standard 3 wave models of stimulated scatter[11] we assume that the laser does not have any effect on LW propagation when explicit SRS coupling is dropped. This propagation is nonlinear with amplitude dependent, isotropic, dispersion relation, $\omega(k,\phi)$. Though in general, $\omega$ is not unique,[12] it is selected as follows[13] by SRS. The SRS daughter LW is spatially localized in a speckle[14], as is its source $\phi_0$, the SRS beat ponderomotive potential. Electrons are longitudinally trapped but they can escape through the speckle's sides if $\phi \lesssim 1$, at rate $\nu_{\text{escape}} \propto v_e / l_\perp$, dissipating LW energy ("side loss"). Requiring that side loss and $\phi_0$ balance, selects the equilibrium. In the strongly trapped regime, $\nu_{\text{escape}} << \omega_b$, with $\omega_b$ the electron bounce frequency, $\omega_b / \omega_{pe} = k\lambda_D \sqrt{e\phi/Te}$, a BGK mode is approached if a nonlinear plasma resonance, real $\omega$ solutions of $\text{Re}\left[ \varepsilon\left(k, \omega | \phi \right) \right] = 0$, can be found. $\varepsilon$ is the nonlinear dielectric function. This requires small enough $k\lambda_D$ and $\phi < \phi_{\text{LOR}}\left(k\lambda_D\right)$, the loss of resonance (LOR) wave amplitude[13], a kinetic generalization of the warm fluid wave breaking limit.[15] Though these modes are apparently SRS optimal in the sense that they seem to maximize the SRS convective gain rate, $\kappa \propto -\text{Im}\left(1/\varepsilon\right) = \text{Im}\left(\varepsilon\right)/\left|\varepsilon\right|^2 \leq 1/\text{Im}\left(\varepsilon\right)$, it is shown in the next section that the actual optimum deviates from a BGK mode, and the difference becomes noticeable as $\phi_{\text{LOR}}$ is approached from below. We will also show that as $\phi$ crosses $\phi_{\text{LOR}}$ and the BGK mode correspondence is lost, that there is rapid qualitative change in the optimal LW response. Optimal LW response is found by varying $\omega$, for given $k$ and $\phi$, until $\kappa$ is a maximum.[16] Since SRS tends to auto-select the LW response that yields the largest gain, we demand that a model's equilibrium solutions are SRS optimal, our only SRS specific ansatz. The resultant $\omega\left(k, \phi\right)$ and $\text{Im}\left[1/\varepsilon\left(k, \phi\right)\right]$ is a generalized LW dispersion relation since it does not correspond with a true traveling wave solution to Vlasov-Poisson unless $\phi < \phi_{\text{LOR}}\left(k\lambda_D\right)$. In the next section we explore how optimal $\omega$ and $\text{Im}\left(1/\varepsilon\right)$ qualitatively depend on parameters.

## *A.* Optimal Langmuir wave dispersion relation

The hallmark of electron trapping effects on Langmuir waves is reduction of Landau damping, $\gamma_L$. In 1D steady state, every electron under the influence of a traveling electrostatic wave is in either a trapped or passing orbit for all time and strictly 1D Vlasov-Poisson traveling wave solutions enjoy the complete absence of damping[17]. In the approach to this 1D steady state, O'Neil has shown[18] that $\gamma_L \to 0$ on the time scale $1/\omega_b$. This order unity effect contrasts with the associated LW frequency shift, which may be small, varying as[7] $\sqrt{\phi}$. In other words, $\text{Re}(\varepsilon)$ and its nonlinear resonance may change by a small amount, while $\text{Im}(\varepsilon)$ is reduced to zero. In 2D and 3D speckles, electrons escape on the time scale $\sim l_\perp / v_e$, so that $\text{Im}(\varepsilon)$ cannot vanish. These two features suggest a simple model that reveals qualitative implications for optimal SRS theory. Ignore the small nonlinear frequency by setting $\text{Re}(\varepsilon) = \text{Re}(\varepsilon_0)$, and reduce $\text{Im}(\varepsilon)$ (proportional to LW damping) by a factor $r \leq 1$, a constant,

$$\text{Im}(\varepsilon) = r \, \text{Im}(\varepsilon_0). \tag{2}$$

$\varepsilon_0$ is the linear dielectric function for thermal plasma. Fig. 1, shows that as $r$ decreases, the optimal LW frequency, $\omega(k, r)$, approaches linear resonance, $\lim_{r \to 0} \omega(k, r) = \omega_0(k)$, with $\text{Re}\,\varepsilon_0[k, \omega_0(k)] = 0$, if $k \lesssim 0.53$. For larger $k$, resonance is not possible. A natural reference frequency in this regime is $\omega(k, r = 1)$, the linear optimal frequency (dashed curve). It lies below the long-short-dashed top curve, "Landau frequency", the real part of the least damped complex $\omega$ solution of $\varepsilon_0(k, \omega) = 0$. In the deep kinetic regime, $k > 0.53$, a large frequency shift is possible due to damping reduction alone. Loss of resonance at $k \approx 0.53$ is reflected in the optimal response, $-\text{Im}(1/\varepsilon)$, first exceeding linear response (dashed curve in Fig. 2) for $k < 0.53$, followed by its rapid decrease below linear response, as $k$ passes through 0.53, provided that $r << 1$.

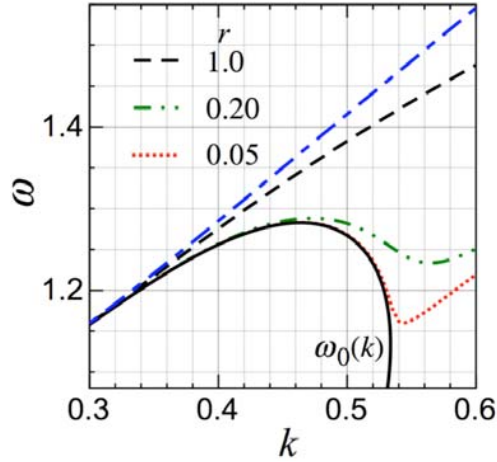

FIG. 1. Simple model SRS optimal LW frequency approaches linear resonance, $\omega_0(k)$ (solid curve), as damping reduction factor, $r$, decreases, if $k \lesssim 0.53$. The long-short-dashed top curve is the Landau frequency.

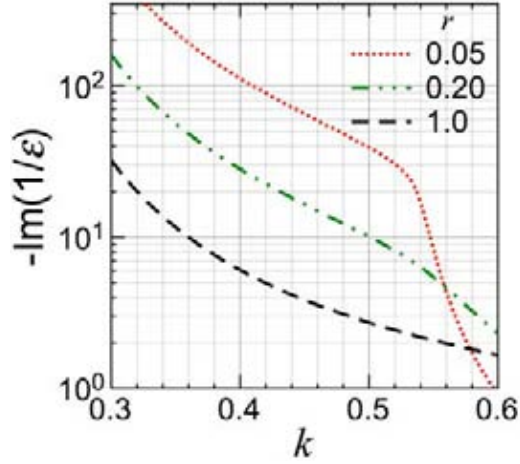

FIG. 2. Simple model optimal LW response sharply decreases, as the LW wavenumber passes loss of resonance, if Landau damping is strongly reduced. Curves again parameterized by damping reduction factor, $r$.

Perturbative analytical results for $\delta\varepsilon$, $\varepsilon = \varepsilon_0 + \delta\varepsilon$ have been obtained[13] (see Eqs. (12) and (13) below) in the $\omega_b / \nu_{escape} >> 1$ limit leading to[19] $r \propto \nu_{escape} / \omega_b$ as well as a perturbation to $\mathrm{Re}(\varepsilon_0)$. This is used in the following. For very small values of $\nu_{escape}$, figures 1 and 2 are qualitatively reproduced with $\phi$ in lieu of $r$ as a parameter. But even merely small values of $\nu_{escape}$, e.g., $\nu_{escape} = O(0.01)\omega_{pe}$, typical of current long laser pulse experiments, are large enough to show richer behavior in optimal LW properties. Fig. 3 shows optimal LW frequency for $\nu_{escape} = 0.01$, and various values of $\phi$, illustrating the effects of nonlinear corrections to $\mathrm{Re}\,\varepsilon_0$, as well as frequency reduction due to reduced damping, the only effect present in Fig. 1. Fig. 4 shows the optimal LW response.

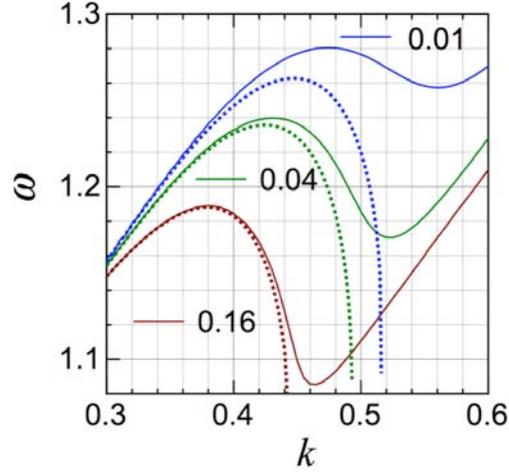

FIG. 3. Optimal LW frequency for $v_{escape} = 0.01\omega_{pe}$, solid curves, labeled by scaled wave amplitude $e\phi/T_e$. The dotted curves, close to and below the small $k$ portion of each corresponding solid curve, are the nonlinear resonances, $\mathrm{Re}\,\varepsilon = 0$.

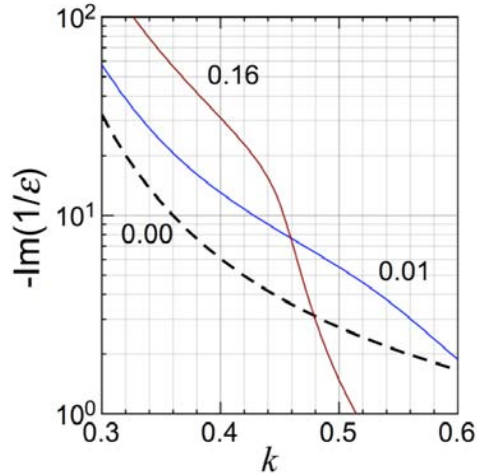

FIG. 4. Optimal LW response, for $v_{escape} = 0.01\omega_{pe}$, with curves labeled by $e\phi/T_e$. Response decreases sharply, for large amplitude waves, as $k$ exceeds loss of nonlinear resonance value.

When a nonlinear LW resonance is possible, Fig. 3 shows that the optimal LW response has a slightly larger than resonant phase velocity, $v=\omega/k$, consistent with the fact that $\mathrm{Im}\,\varepsilon$ decreases as v increases and therefore the optimum response is achieved by allowing finite, but small, $\mathrm{Re}\,\varepsilon$. Except for this correction, SRS optimal theory reduces to previous BGK mode theory[13] when resonance is possible and then $-\text{Im}\left(1/\varepsilon\right) \approx \left[\text{Im}\left(\varepsilon\right)\right]^{-1}$. However, as $k$ (or $\phi$ with $k$ fixed) passes through LOR, the graphs of optimal $\text{Re}\,\varepsilon$ and $\text{Im}\,\varepsilon$ cross, as shown in Fig. 5a for $e\phi/T_e = 1$. For $k$ finitely less than $k_{\text{LOR}} \approx 0.32$, optimal response is near resonance with $\text{Re}\,\varepsilon << \text{Im}\,\varepsilon$. But as $k$ continues to increase past $k_{\text{LOR}}$, $\text{Re}\,\varepsilon$ eventually becomes large compared to $\text{Im}\,\varepsilon$. This implies a somewhat counterintuitive SRS gain rate dependence on $k$ (or equivalently, on $\phi$ at fixed $k$), illustrated in Fig. 5b: the SRS gain rate, $\propto -\text{Im}\left(1/\varepsilon\right)$, decreases by three orders of magnitude as $k$ increases from 0.3 to 0.4 while $\text{Im}\,\varepsilon$ increases by only a factor of 5. Since $|\phi| \propto 1/|\varepsilon|$ (if $1/|\varepsilon| >> 1$) the ratio of the two curves in Fig. 5b is a measure of the relative efficiency with which the laser drives the SRS daughter LW versus the efficiency with which it amplifies backscatter: past LOR a relatively large amplitude LW is needed to result in a given SRS gain.

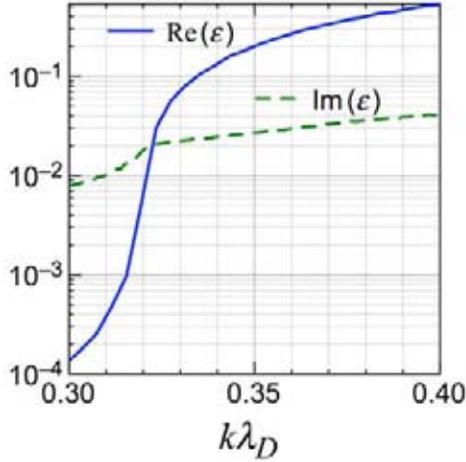    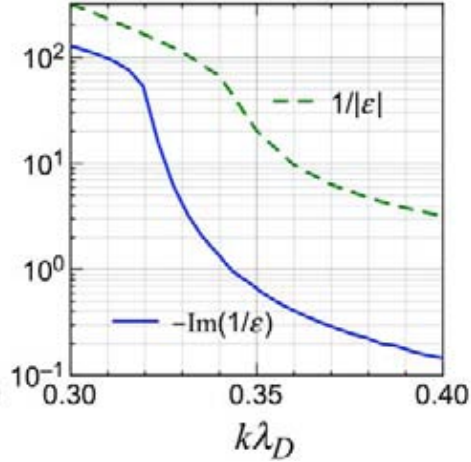

Fig. 5a                            Fig. 5b

FIG. 5a. Optimal Langmuir wave response (for $v_{\text{escape}} = 0.01\omega_{pe}$ and $e\phi/T_e = 1$) has a nonlinear dielectric function, $\varepsilon$, that is nearly resonant for Langmuir wave wavenumber, $k$, less than its loss of resonance value, $k_{\text{LOR}} \approx 0.32$, and increasingly far from resonance as $k$ exceeds $k_{\text{LOR}}$.

FIG. 5b. Optimal Langmuir wave response, for same parameters as previous Fig., shows that past loss of resonance a larger amplitude Langmuir wave is required to attain a given SRS gain rate.

## B. SRS optimal reduced Langmuir wave model and modulational instability

Here it is shown how optimal LW response and its generalized nonlinear dispersion relation, $\omega\left(k,|\phi|,v_{\text{escape}}\right)$, the solid curves in Fig. 3, lead to a reduced model, Eq. (4), for the LW potential envelope, $\phi$, $\Phi = \text{Re}\left(\phi\exp i\mathbf{k}\cdot\mathbf{x}\right)$, in the resonant regime. For slight deviations from equilibrium, $\varepsilon\phi = \phi_0$, modulation theory[20] leads to

$$\frac{\partial\varepsilon}{\partial\omega}\left[i\frac{\partial}{\partial t} + \text{Re}\left(\frac{\varepsilon}{\partial\varepsilon/\partial\omega}\right) + i\,\text{Im}\left(\frac{\varepsilon}{\partial\varepsilon/\partial\omega}\right)\right]\phi = \phi_0. \tag{3}$$

Explicit dependence upon $v_{\text{escape}}$ will usually be suppressed in the following, and if unambiguous by context, $|\phi|$ will more simply be denoted by $\phi$. From this stage forward temporal enveloping is a formality and is omitted for clarity. Assume resonance is possible so that $\partial\varepsilon/\partial\omega$ may be approximated as purely real valued, and demand that Eq. (3) reproduce the optimal LW response to obtain

$$\text{Re}\left(\frac{\partial\varepsilon}{\partial\omega}\right)\left[i\left(\frac{d}{dt} + v_{\text{L}}\right) - \omega\left(k,\phi\right)\right]\phi = \phi_0, \tag{4}$$

and

$$v_{\text{L}}\left(k,\phi\right) = \text{Im}\left(\varepsilon\right)/\text{Re}\left(\partial\varepsilon/\partial\omega\right). \tag{5}$$

In Eq. (5) the factor $\text{Im}\left(\varepsilon\right)$ is related to the optimal LW response by $\text{Im}\left(\varepsilon\right) = -1/\text{Im}\left(1/\varepsilon\right)$, (see, $e.g.$, Fig. 4). If optimal LW response is assumed nearly resonant, then

$$\text{Re}\left\{\varepsilon\left[k,\omega\left(k,\phi\right),\phi\right]\right\} \approx 0, \tag{6}$$

and $\partial\varepsilon/\partial\omega$ may be evaluated implicitly from this resonance condition. Optimal LW theory gives $\omega\left(k,\phi\right)$ and response, but modulation theory is required to connect with generalized Landau damping, $v_{\text{L}}$, given by Eq. (5). This approach breaks down as LOR is approached since $\mathrm{Re}\left(\partial\varepsilon/\partial\omega\right)\to 0$ implying that higher order time derivative terms must be retained, a regime beyond the scope of the remainder of this paper.

Even if $\varepsilon$ is evaluated perturbatively, $\omega\left(k,\phi\right)$ as determined by Eq. (6) contains terms to all order in $\phi$ and, while not exact, is non-perturbative. The difference between this result and straight perturbative evaluation of the LW frequency shift is illustrated in Fig. 4 of Ref. 8.

### 1. Non-perturbative LW frequency difference and instability

Let $v_{\mathrm{escape}}/\omega_{\mathrm{b}}<<1$, so that $\nu_{\mathrm{L}}$ may be momentarily ignored and equilibrium solutions to Eq. (4) are possible for $\phi_0=0$. Let $\phi\sim\exp i\delta\mathbf{k}\cdot\mathbf{x}$. Since LW propagation is assumed not to depend on direction, $\omega$ only depends on $k=|\mathbf{k}|$, and Eq. (4) therefore implies

$$i\,d\phi/dt=\left[\omega\left(k,\phi\right)+d\omega\left(|\mathbf{k}+\delta\mathbf{k}|,k,\phi\right)\right]\phi\,, \tag{7}$$

$$d\omega\left(|\mathbf{k}+\delta\mathbf{k}|,k,\phi\right)=\omega\left(|\mathbf{k}+\delta\mathbf{k}|,|\phi|\right)-\omega\left(k,\phi\right). \tag{8}$$

The Taylor series expansion of $d\omega$ in $\delta\mathbf{k}$ yields group velocity and diffraction terms while higher order derivative terms are usually ignored. It is not necessary, however, to perform the Taylor series expansion. To test for instability, linearize Eq. (8) about equilibrium, $\phi_{\mathrm{eq}}$, $\phi=\phi_{\mathrm{eq}}(t)+\delta\phi(t)\cos\delta\mathbf{k}\cdot\mathbf{x}$, and use $\delta\psi=\delta\phi\exp\left[i\omega\left(k,\phi_{\mathrm{eq}}\right)t\right]$ and its complex conjugate as dependent variables,

$$i\frac{d\delta\psi}{dt}=\mathbf{D}\left(\mathbf{k},\phi_{\mathrm{eq}},\delta\mathbf{k}\right)\delta\psi+\phi_{\mathrm{eq}}\frac{\partial\omega}{\partial|\phi|}\bigg|_{\mathrm{eq}}\delta|\phi|\,. \tag{9}$$

$d\omega$ has been replaced by $D$,

$$2D\left(\mathbf{k},\phi,\delta\mathbf{k}\right)=\omega\left(|\mathbf{k}+\delta\mathbf{k}|,\phi\right)+\omega\left(|\mathbf{k}-\delta\mathbf{k}|,\phi\right)-2\omega\left(k,\phi\right), \tag{10}$$

its even part, because the odd part of $d\omega$ contributes to a nonlocal group velocity term which has no effect on linear instability growth. $\delta|\phi|=0.5\left(\phi_{\mathrm{eq}}\delta\psi^*+\phi_{\mathrm{eq}}^*\delta\psi\right)/|\phi_{\mathrm{eq}}|$, and $\phi_{\mathrm{eq}}$ is now time independent. There is no term of the form $\delta|\phi|\left(\partial D/\partial|\phi|\right)\phi_{\mathrm{eq}}$ in Eq. (8)

because $D$ is a generalized differential operator which when applied to spatially uniform $\phi_{eq}$ gives zero. These formal manipulations miss a subtle but important difference between nonlinear and linearized solutions. Imposing Hamiltonian dynamics results in a quantitatively different model, discussed in section III.B.

Eq. (9) and its complex conjugate have solutions which vary as $\exp(\gamma t)$ with

$$\left(\gamma + \nu_L\right)^2 = -D\left[\phi\frac{\partial\omega}{\partial\phi} + D\right].\tag{11}$$

Restored Landau damping provides an instability cutoff. It and $\partial\omega/\partial\phi$ are evaluated at $(k,\phi)$. This dispersion relation is non-perturbative in $\phi$ and $\delta k$. Instability, $\mathrm{Re}\,\gamma > 0$, is not possible unless $D\partial\omega/\partial\phi < 0$. Perturbative[7], and non-perturbative LW frequency shift evaluation, as shown in Fig. 3, imply that $\partial\omega/\partial\phi < 0$, and thus a dispersive instability cutoff once $D < 0$, and another at $D = -\phi\partial\omega/\partial\phi$. If $\delta k$ were small, instability would be called modulational. Since $\delta k$ is not necessarily small and $\omega$ is evaluated non-perturbatively, this analysis may be viewed as a generalization of previous work[8] on the trapped particle modulational instability, or TPMI. $\gamma$ may be as large as $\gamma_{max} = -\nu_L - 0.5\phi\partial\omega/\partial\phi$ if $D = -0.5\phi\partial\omega/\partial\phi$ is attained for some $\delta k$.

## 2. Filamentation linear instability

Let $\mathbf{k} = (0,k_\parallel)$ and $\delta\mathbf{k} = (k_\perp,\delta k_\parallel)$. Since $2D_\parallel = \partial^2\omega/\partial k_\parallel^2 < 0$ for $k \gtrsim 0.3$, as shown in Fig. 3, and quickly exceeds $2D_\perp = \partial^2\omega/\partial k_\perp^2$ in magnitude (see Fig. 7a and 7b below), TPMI requires finite $k_\perp$ and small $\delta k_\parallel$. For simplicity we here restrict our attention to $\delta k_\parallel = 0$, the filamentation regime. Rotational invariance implies $D_\perp = \mathrm{v}_g/(2k_\parallel)$, with $\mathrm{v}_g = \partial\omega/\partial k_\parallel$ the LW group velocity. Long wavelength analysis, however, which expands $D$ only to 2$^{nd}$ order in $\delta\mathbf{k}$ may fail if $D_\perp$ goes through zero at finite $\phi = \phi_D < \phi_{LOR}$, as is apparently the case in Fig. 3, thus presenting a singular perturbation to a model based solely upon $v_g$ and $D_\perp$. This must be regularized, even in a bare bones model, by inclusion of hyper-diffraction because once $\phi(\mathbf{x}) > \phi_D$ somewhere in plasma, because its omission would lead to solutions with spurious internal boundary layers along the surface $\phi(\mathbf{x}) = \phi_D$. Hyper-diffraction refers to quartic $\delta k$ terms in $D$'s expansion. But first we proceed without expansion in $\delta \mathbf{k}$.

For our examples, $\omega_b / v_{escape}$ is large enough, and $\phi$ small enough, that perturbative results for $\delta \varepsilon$ ( $\varepsilon = \varepsilon_0 + \delta \varepsilon$ ) are useful. We recall from Ref. [13] side loss "Krook" model,

$$k^2 \operatorname{Re}(\varepsilon) \approx k^2 \operatorname{Re}(\varepsilon_0) + 1.76 f_0''(v)\sqrt{\phi} , \tag{12}$$

and

$$k^2 \operatorname{Im}(\varepsilon) \approx -6.2 f_0'(v) v_{escape}/\omega_b \approx 2k^2 \operatorname{Im}(\varepsilon_0) v_{escape}/\omega_b \tag{13}$$

The residual damping component of $\operatorname{Im}(\varepsilon)$, a contribution which is also proportional to $v_{escape}$ but does not vanish as $\phi$ continues to increase, has been omitted because multi-dimensional transit time damping (TTD) analysis of a LW wavepacket shows that damping is overestimated by the Krook model unless $\omega_b / v_{escape} >> 1$: for $\omega_b / v_{escape}$ order unity, enhancement of Landau damping is order $v_{escape}^2$ in TDD[21] but order $v_{escape}$ in the Krook model. In addition, it is found that TPMI stabilizes before loss of resonance is closely approached, allowing approximate LW dispersion determination by resonance, $\operatorname{Re}(\varepsilon) = 0$, in lieu of optimization. This allows a simpler and far more accurate determination of $\partial \omega / \partial \phi$, simply by differentiating $\operatorname{Re}(\varepsilon)$, given, *e.g.*, by Eq. (12), compared with numerically differentiating the numerically determined LW frequency.

The filamentation growth rate determined by Eqs. (5), (6), (10), (11), (12) and (13), is shown for $k = 0.35$ and $v_{escape} = 0.0$ in Fig. 06 as a function of LW amplitude, $\phi$, and fluctuation transverse wavenumber, $k_\perp$.

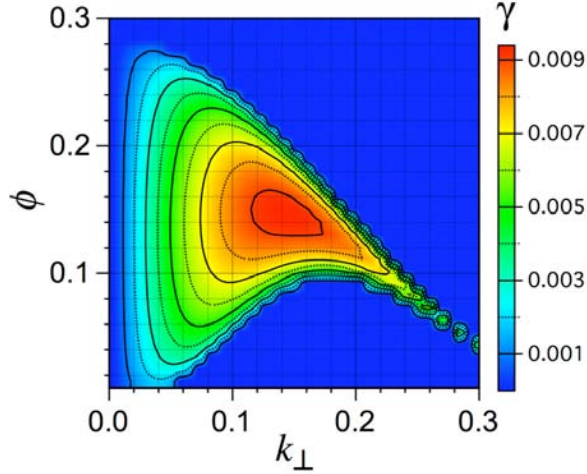

Fig. 6

FIG. 06 Scaled Langmuir wave filamentation instability growth rate, $\gamma/\omega_{pe}$, for fluctuation with (scaled) wavenumber $k_\perp \lambda_D$, has nonstandard contour shape. The family of equilibria has wavenumber $k_\parallel \lambda_D = 0.35$, and variable amplitude, $e\phi/T_e$. Trapped electrons cannot escape ($v_{escape} = 0.0$).

## III. PRACTICAL LW DIFFRACTION MODEL

The filamentation unstable domain, illustrated in Fig. 06, is unusual compared to models with constant diffraction coefficient, $D_\perp$, which have the property that as $-\phi\,\partial\omega/\partial\phi$ increases, so do the domain width in $k_\perp$ and the value of $k_\perp$ at maximum growth rate. This leads to a stability boundary with positive slope while the actual LW upper stability boundary has negative slope, going roughly from $(k_\perp, \phi) = (0, 0.3)$ to $(0.3, 0)$ in Fig. 06, near the graph of $D(k_\parallel, \phi, k_\perp) = 0$, and the lower boundary is near the graph of $D = -0.5\phi\,\partial\omega/\partial\phi$, for $v_{escape} << 1$.

Instability is not possible once $D_\perp$ goes negative. This occurs at $\phi \approx 0.3$ when $k_\parallel = 0.35$, as seen in Fig. 06 and 07b (and Fig.14 of Ref. [8]). As a result, stopping at $2^{nd}$ order in the $k_\perp$ expansion of the generalized LW frequency difference, $D = D_\perp\left(k_\parallel, \phi\right)k_\perp^2$, is disastrous even for linear instability because as $\phi \to 0.3$ and $D_\perp \to 0$, maximum growth rate remains finite but is attained as $k_\perp \to \infty$. In this section stability properties of the hyper-diffraction model are shown to agree qualitatively with results of the previous section, which may be considered consequences of a model that retains terms to all orders in the $k_\perp$ expansion of $D$.

## A. Hyper-diffraction

Rotational invariance, already used to relate $D_\perp$ to $v_g$, also implies

$$4D_{\text{hyper}} = \left(D_\parallel - D_\perp\right)/k_\parallel^2 \,. \tag{14}$$

$D_{\text{hyper}}$ is the hyper-diffraction coefficient, so that to 4[th] order in $k_\perp$,

$$D = D_\parallel\left(\delta k_\parallel\right)^2 + D_\perp k_\perp^2 + D_{\text{hyper}}k_\perp^4 \,. \tag{15}$$

Thus, the coefficients are simply expressible in terms of the slope and curvature of the basic LW dispersion, $\omega\left(k, \phi\right)$. The solid curves in Figs. 07a and 07b shows the graphs of $D_\parallel$ and $D_\perp$ for $k_\parallel = 0.35$, based on the resonant dispersion relation with $\text{Re}\left(\varepsilon\right)$ given by Eq. (12). One need not directly evaluate $k$ derivatives of $\omega$, a method prone to inaccuracy when the dispersion relation is obtained as numerical solution of $\text{Re}\left(\varepsilon\right) = 0$. Instead, evaluate $dk/dv$ and $d^2k/dv^2$ by successively differentiating $\text{Re}\left(\varepsilon\right) = 0$, and then use $d^2v/dk^2 = -\left(dv/dk\right)^3 d^2k/dv^2$, which follows from $\left(dk/dv\right)\left(dv/dk\right) = 1$, and $d\omega/dk = v + k\,dv/dk$, $d^2\omega/dk^2 = 2\,dv/dk + k\,d^2v/dk^2$, which follow from $\omega = kv$, to evaluate $d\omega/dk$ and $d^2\omega/dk^2$ (at constant $\phi$). Solid curve in Fig. 07a illustrates the graph of $D_\parallel\left(k_\parallel = 0.35, \phi\right)$, and in Fig. 07b the graph of $D_\perp\left(k_\parallel = 0.35, \phi\right)$.

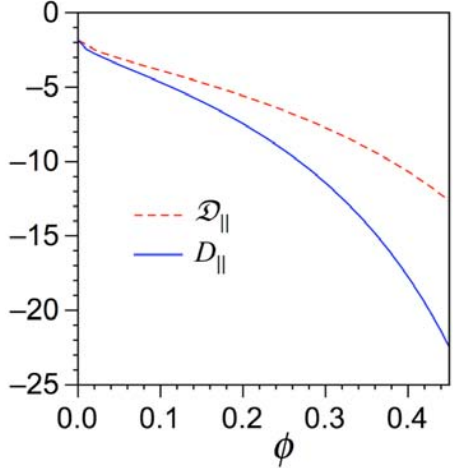
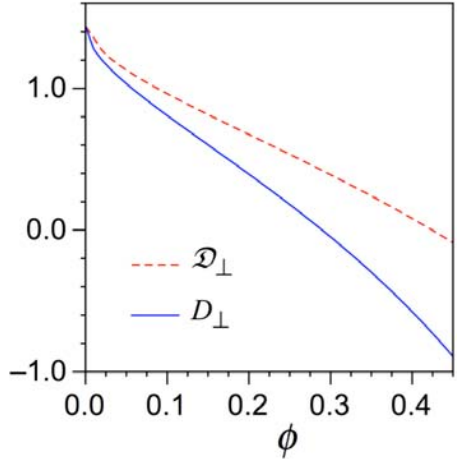

Fig. 07a                    Fig. 07b

FIG. 7a Langmuir wave scaled parallel diffraction coefficient, $D_\parallel$, for LW with $k_\parallel \lambda_D = 0.35$, as a function of LW scaled amplitude, $e\phi/T_e$. Note: for $k_\parallel = \phi = 0$, $D_\parallel = D_\perp = 3/2$.

FIG. 7b Langmuir wave scaled perpendicular diffraction coefficient, $D_\perp$, with $k_\parallel \lambda_D = 0.35$. Dashed curves are corresponding diffraction coefficients for Hamiltonian model, Sec. III.B.

Fig. 08 compares contours (solid curves) of LW frequency difference, $D$, with its hyper-diffraction approximation (dashed curves), Eq. (15), for $k_\parallel = 0.35$ and $\delta k_\parallel = 0$.

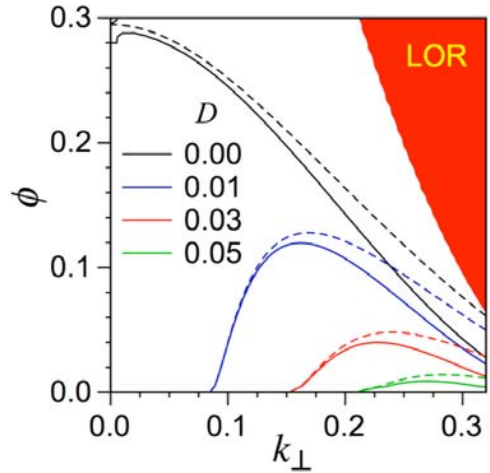

FIG. 08 Contours of Langmuir wave frequency difference (see Eq. (10)), $D$, as calculated from dispersion relation (solid curves) and its hyper-diffraction approximation, Eq. (15), with $k_\parallel = 0.35$ and $\delta k_\parallel = 0$, agree qualitatively away from loss of resonance region.

The two sets of contours agree qualitatively, in particular the $D = 0$ contours agree (this contour sets the filamentation stability boundary for small damping), until the LOR boundary is approached, where frequency derivatives diverge and agreement must falter.

## B. HAMILTONIAN STRUCTURE

A naïve reversion, from Fourier to "x-space", of LW envelope dynamics as given by the hyper-diffraction model (ignoring Landau damping) yields

$$i\left(\frac{\partial}{\partial t} + v_g \frac{\partial}{\partial x_\parallel}\right)\phi = \left[\omega - D_\parallel \frac{\partial^2}{\partial x_\parallel^2} - D_\perp \nabla^2 + D_{hyper}\nabla^4\right]\phi, \quad \nabla^2 = \frac{\partial}{\partial \mathbf{x}_\perp} \cdot \frac{\partial}{\partial \mathbf{x}_\perp}. \tag{16}$$

If the transport coefficients ($v_g$, $D_\parallel$,…) depend on $|\phi|$, then this is not derivable from a Hamiltonian, $H$, *i.e.*, the dynamics is not of the form $i\partial\phi/\partial t = \delta H/\delta\phi^*$, an unsatisfactory feature given that the underlying Vlasov-Poisson dynamics is[22] Hamiltonian. We require that deviations from Hamiltonian dyamics be explicitly linked to electron heating effects. For example, transit time damping in the LW geometry induced by speckle SRS implies a damping rate proportional to the inverse speckle width, $l_\perp^{-1}$, in the strongly trapped regime. However, if the LW phase varies over a distance significantly smaller than $l_\perp$, then the rate of loss of trapped electrons, and consequently plasma heating, increase. Such variation is observed[10] in 2D simulations that manifest LW self-focusing. Another possible source of dissipation is nonlinear $v_g$, which would typically propel smooth initial conditions into singular shocks were it not for $D_\parallel$, but may instead lead to dispersive shocks. If electrons passing through a dispersive shock are significantly heated, then a corresponding dissipation term must be added to the model LW dynamics. One reason for explicitly separating out LW dissipative dynamics is that evolution due to the Hamiltonian alone leaves $H$ invariant, and thus provides a nontrivial test for numerical integration since it depends on high order spatial derivatives. Also, since $H$, as presented below, has no explicit dependence on spatial coordinates, momentum is conserved. More importantly, for extension of this model to include SRS, $H$ is invariant under $\phi \to \phi e^{i\theta}$, $\theta$ constant, implying conservation of plasmon number, $N = \int |\phi|^2 \, d\mathbf{x}$, without which the Manley-Rowe[23] relation could not be recovered when $\phi$ is coupled to the laser and scattered light fields.

A simple Hamiltonian consistent with hyper-diffraction is given by

$$H = \int \left( I_\omega + \mathfrak{v}j + \mathcal{D}_\parallel \left| \partial \phi / \partial x_\parallel \right|^2 + \mathcal{D}_\perp \left| \nabla \phi \right|^2 + \mathcal{D}_{\text{hyper}} \left| \nabla^2 \phi \right|^2 \right) d\mathbf{x}. \qquad (17)$$

Its terms are defined and motivated as follows. The $I_\omega \left( |\phi| \right)$ contribution to $H$ leads to the $\omega\phi$ term in the equation of motion if

$$dI_\omega / d\phi = 2\phi\omega\left( \phi \right). \qquad (18)$$

To lighten notation, here and in the following, $\phi$ is understood to mean $|\phi|$ when the context is clear. LW convection is recovered by coupling the parallel LW flux, $j = \text{Im}\left( \phi^* \partial \phi / \partial x_\parallel \right)$, to the generalized group velocity, $\mathfrak{v}$, a yet to be determined function of $|\phi|$. Eqs. (17) and Eq. (18), imply

$$i\left[ \frac{\partial}{\partial t} + \mathfrak{v}\frac{\partial \phi}{\partial x_\parallel} + \frac{1}{2}\phi\frac{\partial \mathfrak{v}}{\partial x_\parallel} + i\frac{1}{2|\phi|}\frac{d\mathfrak{v}}{d|\phi|}j\phi \right] = (\omega + \delta\omega)\phi - \frac{\partial}{\partial x_\parallel}\left( \mathcal{D}_\parallel \frac{\partial \phi}{\partial x_\parallel} \right) \\ -\nabla \cdot \left( \mathcal{D}_\perp \nabla \phi \right) + \nabla^2 \left( \mathcal{D}_{\text{hyper}} \nabla^2 \phi \right) \qquad (19)$$

Eq. (19) manifestly conserves $N$ for any choice of $\mathfrak{v}\left( |\phi| \right)$. The gradient dependent frequency shift, $\delta\omega$, is a consequence of nonlinear diffraction,

$$\delta\omega = \frac{1}{2|\phi|}\left( \frac{d\mathcal{D}_\parallel}{d|\phi|}\left| \frac{\partial \phi}{\partial x_\parallel} \right|^2 + \frac{d\mathcal{D}_\perp}{d|\phi|}\left| \nabla \phi \right|^2 + \frac{d\mathcal{D}_{\text{hyper}}}{d|\phi|}\left| \nabla^2 \phi \right|^2 \right). \qquad (20)$$

Eq. (19) may be simplified by use of the identity

$$\phi\frac{\partial \mathfrak{v}}{\partial x_\parallel} + i\frac{1}{|\phi|}\frac{d\mathfrak{v}}{d|\phi|}j\phi = |\phi|\frac{d\mathfrak{v}}{d|\phi|}\frac{\partial \phi}{\partial x_\parallel},$$

to obtain

$$i\left[\frac{\partial}{\partial t} + \left(\mathfrak{v} + \frac{1}{2}|\phi|\frac{d\mathfrak{v}}{d|\phi|}\right)\frac{\partial}{\partial x_{\parallel}}\right]\phi = \left(\omega + \delta\omega\right)\phi$$
$$-\frac{\partial}{\partial x_{\parallel}}\left(\mathfrak{D}_{\parallel}\frac{\partial\phi}{\partial x_{\parallel}}\right) - \nabla\cdot\left(\mathfrak{D}_{\perp}\nabla\phi\right) + \nabla^2\left(\mathfrak{D}_{\text{hyper}}\nabla^2\phi\right) \tag{21}$$

Note that when Eq. (21) is linearized about a spatially uniform solution, $\delta\omega$, being 2$^{\text{nd}}$ order in the fluctuation, does not affect modulational instability analysis. By definition of $\mathfrak{v}_g$ and $D_{\parallel}$, initial condition $\phi = |\phi|\exp\left(i\delta k x_{\parallel}\right)$ evolves as

$$i\partial\phi/\partial t = \left[\omega + \mathfrak{v}_g\delta k + D_{\parallel}\left(\delta k\right)^2 + \ldots\right]\phi. \tag{22}$$

Consistency of Eq. (21) with Eq. (22) to first order in $\delta k$ requires

$$\mathfrak{v} + 0.5\phi \, d\mathfrak{v}/d\phi = \mathfrak{v}_g. \tag{23}$$

Consistency with Eq. (22) to 2$^{\text{nd}}$ order in $\delta k$, similarly requires

$$\mathfrak{D}_{\parallel} + 0.5\phi \, d\mathfrak{D}_{\parallel}/d\phi = D_{\parallel}. \tag{24}$$

Eqs. (23) and (24) have the solutions

$$\phi^2\mathfrak{v} = 2\int_0^{\phi}\phi\mathfrak{v}_g\mathrm{d}\phi, \tag{25}$$

$$\phi^2\mathfrak{D}_{\parallel} = 2\int_0^{\phi}\phi D_{\parallel}\mathrm{d}\phi. \tag{26}$$

Rotational invariance implies $2\mathfrak{D}_{\perp} = \mathfrak{v}/k$ and $4\mathfrak{D}_{\text{hyper}} = \left(\mathfrak{D}_{\parallel} - \mathfrak{D}_{\perp}\right)/k^2$. Since $\mathfrak{D}_{\perp}$ is the $\phi^2$ average of $D_{\perp}$, and $D_{\perp}$ is a monotonically decreasing function of $\phi$, $\mathfrak{D}_{\perp} > D_{\perp}$, as seen in Fig. 07b. Eq. (22), and $\mathfrak{D}_{\perp} \propto \mathfrak{v}$, therefore imply $d\mathfrak{D}_{\perp}/d\phi < 0$. Similarly, since $D_{\parallel}$ is also a monotonically decreasing function of $\phi$, $d\mathfrak{D}_{\parallel}/d\phi < 0$. If $\mathfrak{D}_{\text{hyper}}$ also monotonically decreased, as suggested by analysis of the data shown in Fig. 7, then $d\mathfrak{D}_{\text{hyper}}/d\phi < 0$, implying $\delta\omega \leq 0$ which would reinforce $\omega < 0$.

Recall that qualitative agreement of the particular truncation, Eq. (15), with exact $D$, has been demonstrated in Fig. 08. If for purposes of this discussion, the distinction between the two is ignored, then substitution of $D = D_{\parallel}\left(\delta k_{\parallel}\right)^2 + D_{\perp}k_{\perp}^2 + D_{\text{hyper}}k_{\perp}^4$ into Eq. (11) regains the original TPMI growth rate. However, the Hamiltonian constrained diffraction and hyper-diffraction coefficients, $\mathfrak{D}_{\|}$, $\mathfrak{D}_{\perp}$ and $\mathfrak{D}_{\text{hyper}}$, are related to but different than those defined as $k$ derivatives of $\omega$. If instead of Eq. (15),

$$D = \mathfrak{D}_{\|}\left(\delta k_{\|}\right)^2 + \mathfrak{D}_{\perp}k_{\perp}^2 + \mathfrak{D}_{\text{hyper}}k_{\perp}^4, \qquad (27)$$

is substituted into, Eq. (11), as follows from linearized Eq. (21), the resultant filamentation growth rate is shown in Fig. 09a for the same parameters as in Fig. 06. The unstable region has expanded compared to that shown in Fig. 06, and the maximum growth rate has increased by about $1/3^{\text{rd}}$. The case $v_{\text{escape}}/\omega_{pe} = 0.005$ is shown in Fig. 09b.

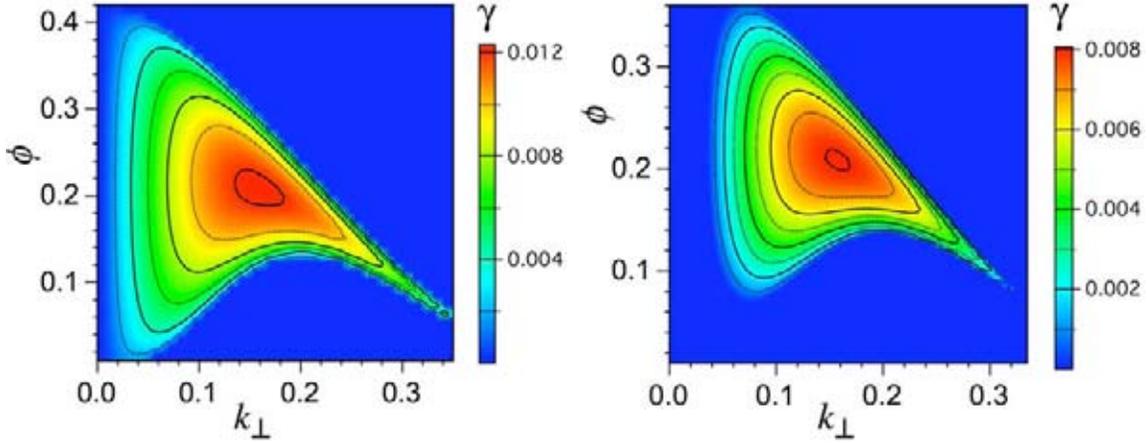

Fig. 09a                              Fig. 09b

FIG. 09a. Langmuir wave filamentation instability growth rate for $k_{\|}\lambda_D = 0.35$ and $v_{\text{escape}} = 0.0$, based upon Hamiltonian constrained diffraction, Eq. (27).

FIG. 09b. Langmuir wave filamentation instability growth rate is shown for same value of $k_{\|}$ but $v_{\text{escape}}/\omega_{pe} = 0.005$.

Practically, one may not want to deal with terms beyond hyper-diffraction, but to any finite order in $\delta\mathbf{k}$, the relationship between the coefficient in the expansion of $D$, and those required for compatibility with Hamiltonian dynamics, Eq. (21), is the same as illustrated by Eq. (26). One could therefore calculate, with $D$ given by Eq. (10),

$$\phi^2 \mathfrak{D} = 2\int_0^\phi \phi D d\phi . \qquad (28)$$

Substituting $\mathcal{D}$ for $D$ in Eq. (11) yields an expression for the TPMI growth rate that is non-perturbative in $\delta\mathbf{k}$ and $\phi$, and compatible with Hamiltonian dynamics. This generalization is not pursued here.

## IV. COMPARISON WITH SIMULATION

2D PIC simulation of the Vlasov-Poisson system, with SRS relevant BGK mode initial conditions in 2D, has been performed, as well as 2D PIC simulation of the Vlasov-Maxwell system, with diffraction limited light wave beam boundary condition of sufficient intensity to yield roughly 1% average backscatter SRS [10]. The former initially had 1D electric field variation except for particle noise, $k_\parallel \lambda_D = 0.35$ and amplitude in the range $0.15 < e\phi/T_e < 0.5$, as discussed in Fig. 9 of Ref. [10]. Its boundary conditions were doubly periodic, 40 LW wavelengths long in the perpendicular direction, and only one along $x_\parallel$. Clear exponential growth of the $\left|\phi\left(k_\perp \lambda_D = 0.05\right)\right|$ mode was observed for larger values of $\phi$, but except for growth rate magnitude, $O\left(0.01\omega_{pe}\right)$, which qualitatively agrees with that predicted in Fig. 09a, agreement is not good. For example, simulation growth rates increase with $\phi$ in the range $0.15 < e\phi/T_e < 0.5$, in particular, $\gamma \approx 0.015\omega_{pe}$ when $e\phi/T_e = 0.5$, while $e\phi/T_e \approx 0.4$ is at the predicted instability region boundary. This simulation study should be expanded to determine the maximum value of $\phi$ for which instability is possible at any value of $k_\perp \lambda_D$, and for smaller values of $\phi$, determine $k_\perp \lambda_D$ at maximum growth rate how large can $k_\perp \lambda_D$ be before instability is lost.

The SRS simulations employed outgoing wave boundary conditions, and electrons leaving the system are replenished by sampling the initial background thermal distribution. Comparison with linear instability theory developed here is tenuous because of the intrinsically non-equilibrium field configurations actually sampled by the simulation. We refer the reader again to Ref. [10] for detailed discussion, except to recall a key qualitative feature of that work: SRS daughter LW filamentation was observed for $0.19 \leq k_\parallel \lambda_D \leq 0.45$ but not yet tested at higher values of $k_\parallel \lambda_D$. Since $D_\perp \left( k\lambda_D, \phi = 0 \right) < 0$ for $k\lambda_D > 0.45$ (see Fig. 8 of Ref. [8]), theory predicts no chance of linear filamentation instability at the largest $k_\parallel \lambda_D$.

# V. DISCUSSION

It has been argued that stimulated Raman scatter (SRS) relevant Langmuir wave (LW) equilibria are optimal in the sense that they maximize the SRS convective gain rate. From this constraint, along with standard requirements of rotational invariance and Hamiltonian dynamics, a Langmuir wave dispersion relation emerges, $\omega\left( k, \phi, v_{\text{escape}} \right)$, for given nonlinear dielectric function, $\varepsilon$. The dispersion relation depends on $\phi$, the wave's electrostatic potential amplitude, its wavenumber, $k$, and $v_{\text{escape}}$, the rate at which trapped electrons are exchanged with a statistical sample of the background plasma distribution function, $f_0$. Specific calculations presented here assume thermal $f_0$, as appropriate for SRS originating in a single laser speckle. For small enough $k$, $\phi$ and $v_{\text{escape}} / \left( k v_e \sqrt{e\phi / T_e} \right)$, the optimal LW is nearly resonant, $\text{Re}\left( \varepsilon \right) \approx 0$, and thus close to a BGK mode selected by equilibration of the LW under the influence of a weak beat ponderomotive source and slow dissipation at a rate $\propto v_{\text{escape}}$.

This leads to a LW model in which the electrostatic potential's spatial envelope, $\phi$, satisifies Eq. (21), absent dissipation. The transport coefficients of this model are related to the usual group velocity, $v_g = \partial\omega / \partial\phi$, and diffraction coefficients as illustrated by Eqs. (23) and (24). Linearization leads to a filamentation instability with non-standard instability domain, Fig. 09.

Dissipation depends on LW geometry, which is highly variable[10] during SRS evolution. For example LW self-focusing, the nonlinear stage of the filamentation instability, generates spatial scales small compared to a laser speckle width, thereby increasing $v_{escape}$ from its trapping reduced value. Simulations[10] suggest that this is the dominant backscatter SRS saturation mechanism in the intermediate kinetic regime, roughly $0.20 < k\lambda_D < 0.45$.

The LW model, Eq. (21), has the standard group velocity, $v_g$, term when Eq. (23) is applied. Electron trapping leads to its dependence on LW amplitude, as is apparent by inspection of Fig. 03, and would typically lead to a parallel (that is, along the LW envelope wavevector direction, $\mathbf{k}/k$) shock unless parallel diffraction, $\mathfrak{D}_\parallel$, is included in the wave dynamics. For large enough $\phi$, $v_g$ goes negative, and rotational invariance implies the same for $\mathfrak{D}_\perp$. This provides a filamentation instability cutoff since instability requires $\mathfrak{D}_\perp > 0$. However, unless hyper-diffraction is included, an unphysical singular layer would develop on the surface $\mathfrak{D}_\perp(\mathbf{x}) = 0$. Thus Eq. (21) is the simplest possible model consistent with these basic LW properties.

An even simpler mode is suitable if $\phi$ remains small enough, so that $\mathfrak{D}_\perp$ is bounded away from zero, and then $\mathfrak{D}_{hyper}$ could be ignored without risk of $\mathfrak{D}_\perp$ becoming a singular perturbation. If one makes the additional ansatz that the group velocity nonlinearity is ignorable, then $v_g$ and $\mathfrak{D}_\perp$ may be taken as positive constants, allowing the much simpler model obtained by also setting $\mathfrak{D}_\parallel = 0$. This may well be a suitable starting point for understanding the properties of Eq. (21) even if the only nonlinearities retained are the trapped electron LW frequency shift and damping reduction, and, to provide a source of energy, SRS coupling. Since solutions of even this greatly simplified model are expected to exhibit rich space-time behavior, such as SRS onset, LW bowing and self-focusing[10], with dependence on multiple dimensionless parameters, including $k\lambda_D$, $v_{escape}/\omega_{pe}$, scaled system length, scaled SRS growth rate and spatial dimension, it may prove a useful starting point. If $k\lambda_D \lesssim 0.30$, then the Langmuir wave ion-acoustic decay instability may beat[8][24][25] Langmuir wave self-focusing if scaled acoustic wave damping rate is not too large, introducing at least one more parameter.

[1] Y. Kato et al., "Random Phasing of High-Power Lasers for Uniform Target Acceleration and Plasma-Instability Suppression," *Physical Review Letters* **53**, no. 11, 1057 (1984).

[2] Harvey A. Rose and D. F. DuBois, "Laser hot spots and the breakdown of linear instability theory with application to stimulated Brillouin scattering," *Physical Review Letters* **72**, no. 18, 2883 (1994).

[3] Philippe Mounaix and Laurent Divol, "Near-Threshold Reflectivity Fluctuations in the Independent-Convective-Hot-Spot-Model Limit of a Spatially Smoothed Laser Beam," *Physical Review Letters* **89**, no. 16 (2002), 165005.

[4] Ira B. Bernstein, John M. Greene, and Martin D. Kruskal, "Exact Nonlinear Plasma Oscillations," *Physical Review* **108**, no. 3, 546 (1957).

[5] Wallace M. Manheimer and Robert W. Flynn, "Formation of Stationary Large Amplitude Waves in Plasmas," Physics of Fluids **14**, 2393(1971).

[6] R. L. Dewar, "Frequency Shift Due to Trapped Particles," *Physics of Fluids* **15**, 712 (1972).

[7] G. J. Morales and T. M. O'Neil, "Nonlinear Frequency Shift of an Electron Plasma Wave," *Physical Review Letters* **28**, 417(1972).

[8] Harvey A. Rose, "Langmuir wave self-focusing versus decay instability," *Physics of Plasmas* **12**, no. 1, 012318 (2005).

[9] In a speckle the ponderomotive expulsion of plasma density by variations of laser intensity may induce a frequency shift in excess of that provided by trapped electrons.

However, this comparison invokes at least two dimensionless parameters, the speckle coherence time (or pulse length) normalized to the ion-acoustic speckle transit time and the electron quiver velocity in the laser EM field normalized to the electron thermal speed. For purposes of this paper, the laser pulse duration is assumed small compared to the ion-acoustic time scale.

[10] L. Yin, B. J. Albright, K. J. Bowers, W. Daughton, and H. A. Rose, "Saturation of backward stimulated Raman scattering of laser from kinetic simulations: Wavefront bowing, trapped particle modulational instability and self-focusing of electron plasma waves", submitted to Physics of Plasmas (2007).

[11] William L. Kruer, *The Physics of Laser Plasma Interactions*, Addison-Wesley, New York, 1988.

[12] For example see Didier Benisti and Laurent Gremillet, "Nonlinear plasma response to a slowly varying electrostatic wave, and application to stimulated Raman scattering," Physics of Plasmas **14**, no. 4, 042304 (2007).

[13] Harvey A. Rose and David A. Russell, "A self-consistent trapping model of driven electron plasma waves and limits on stimulated Raman scatter," *Physics of Plasmas* **8**, no. 11, 4784 (2001).

[14] There are various localization mechanisms including: LW dissipation, ponderomotive depletion of plasma density and trapped electron frequency shift self-localization.

[15] T. P. Coffey, "Breaking of Large Amplitude Plasma Oscillations," *Physics of Fluids* **14**, no. 7, 1402 (1971).

[16] Harvey A. Rose, "Trapped particle bounds on stimulated scatter in the large $k\lambda_D$ regime," *Physics of Plasmas* **10**, no. 5, 1468 (2003).

[17] James Paul Holloway and J. J. Dorning, "Undamped plasma waves," *Physical Review A* **44**, no. 6, 3856 (1991) .

[18] Thomas O'Neil, "Collisionless Damping of Nonlinear Plasma Oscillations," *Physics of Fluids* **8**, no. 12, 2255 (1965).

[19] See Eqs. (48) and (71) of Ref. [13].

[20] Bruce I. Cohen and Allan N. Kaufman, "Nonlinear plasma waves excited near resonance," *Physics of Fluids* **20**, no. 7, 1113 (July 1977).

[21] This true so long as the LW wavepacket amplitude goes to zero smoothly with distance from wavepacket center, otherwise, transit time damping also has a linear correction to Landau damping for small $v_{escape}$ .

[22] Philip J. Morrison, "The Maxwell-Vlasov equations as a continuous Hamiltonian system, Physics Letters A **80**, Issue: 5-6, 383 (1980).

[23] J.M. Manley and H.E. Rowe, "Some General Properties of Nonlinear Elements-Part I. General Energy Relations", Proceedings of the IRE, **44**, Issue 7, 904 (1956).

[24] J. L. Kline et al., "Different $k\lambda_D$ regimes for nonlinear effects on Langmuir waves," *Physics of Plasmas* **13**, no. 5, 055906 (2006).

[25] J. L. Kline et al., "Observation of a Transition from Fluid to Kinetic Nonlinearities for Langmuir Waves Driven by Stimulated Raman Backscatter," *Physical Review Letters*, **94**, 175003 (2005).